\documentclass[aps,prd,preprint,showkeys,showpacs,unsortedaddress,superscriptaddress]{revtex4-1}
\usepackage{amsmath, amsfonts, amssymb, mathrsfs,float}
\usepackage{graphicx}
\usepackage{bm,color}
\usepackage{multirow}
\newcommand{\nn}{\nonumber}

\allowdisplaybreaks

\begin{document}

\title{The second post-Newtonian motion in Reissner-Nordstr\"{o}m spacetime}

\author{Bo Yang}
\affiliation{School of Mathematics and Physics, University of South China, Hengyang, 421001, China}

\author{Chunhua Jiang}
\affiliation{School of Mathematics and Physics, University of South China, Hengyang, 421001, China}

\author{Wenbin Lin}
\email{lwb@usc.edu.cn.}
\affiliation{School of Mathematics and Physics, University of South China, Hengyang, 421001, China}
\affiliation{School of Physical Science and Technology, Southwest Jiaotong University, Chengdu, 610031, China}

\date{\today}

\begin{abstract}
We derived the second post-Newtonian solution for the quasi-Keplerian motion of a charged test particle in the Reissner-Nordstr\"{o}m spacetime under the harmonic coordinates. The solution is formulated in terms of the test particle's orbital energy and angular momentum, both of which are constants at the second post-Newtonian order. The charge effects on the test particle's motion including the orbital period and perihelion precession are displayed explicitly. Our results can be applied to the cases in which the test particle has small charge-to-mass ratio, or the test particle has arbitrary charge-to-mass ratio but the multiplication of the test particle and the gravitational source's charge-to-mass-ratios is much smaller than 1.
\end{abstract}

\maketitle

\section{introduction}

The motion of bodies in the gravitational fields is a classical problem in astronomy and cosmology. For the cases in which the gravitational fields are not extremely strong, the motion can be studied in the post-Newtonian (PN) approximations.
A large number of analytical PN solutions for the motion of the binary systems have been obtained, including the first and higher PN effects of the mass~\cite{Brumberg1972,Soffel1987,Soffel1989,KlionerKopeikin1994,KopeikinEfroimskyKaplan2012,DamourSchafer1988,SchaferWex1993,MGS2004,Boetzel2017,Cho2018,SoffelHan2019}, the 1.5PN effects of the spin-orbit coupling effects~\cite{Wex1995,GPV1998a,GPV1998b,GPV1998c,KonigsdorfferGopakumar2005,KMG2005,KonigsdorfferGopakumar2006b,GopakumarSchafer2011,BMFB2013,GergelyKeresztes2015,Mikoczi2017} on the general motion and the 2PN effects of the mass quadrupole on the circular motion~\cite{Poisson1998}. When taking the limit of extreme mass ratio, the solutions for the motion of the binary systems return to the solution of the motion of the test particle. Based on these studies, we have achieved the
analytical solutions for the motion of the test particle under the most generally parameterized PN force~\cite{YangLin2020a,YangLin2020b}, and those in the spacetime of the classical black holes, including the 2PN effects of the mass in the Wagoner-Will-Epstein-Haugan representation~\cite{YangLin2020d},
the 2PN effects of the spin-induced quadrupole on the equatorial motion in Kerr spacetime~\cite{YangLin2020c} and the 2.5PN effects of the spin-orbit coupling on the general motion in the Kerr spacetime~\cite{YangLin2021}.


The Reissner-Nordstr\"{o}m metric~\cite{Reissner1916,Weyl1917,Nordstrom1918} is the unique spherically symmetric and asymptotically flat solution of the Einstein-Maxwell equations, which describes the exterior spacetime around an isolated spherical object of mass and electric charge. Although the charged astronomical bodies have not been discovered, the possibility of their existence in the universe may not be ruled out. In fact, the charge effects of the astrophysical black holes and stars have been studied extensively~\cite{Iorio2012,ZhaoXie2016,WangShenXie2019,BiniGeralicoRuffini2000,Hossain2000,BiniGeralicoRuffini2007,PuglieseQuevedoRuffini2011b,PuglieseQuevedoRuffini2017,YangLin2019b,BakryMoatimidTantawy2021}. For example, Bini, Geralico and Ruffini have investigated the circular motion of a charged test particle in the field of the Reissner-Nordstr\"{o}m black hole~\cite{PuglieseQuevedoRuffini2011b,PuglieseQuevedoRuffini2017}. In our previous work, we have shown that the black hole's electric charge can contribute to the orbital perihelion precession at the 1PN order, but not to the orbital period at the same order~\cite{YangLin2019b}. It is interesting to further explore the effects of the black hole's charge on the particle's motion including the orbital perihelion precession and period at the higher PN orders.
On the other hand, the electrically charged test particles are all around in astrophysics, so it is especially important to discuss the motion of the charged test particles in the field of the Reissner-Nordstr\"{o}m black hole.

In this work, we derive the 2PN solution for the quasi-Keplerian motion of the charged test particle in the Reissner-Nordstr\"{o}m spacetime, and exhibit the charge effects on the orbital period and perihelion precession. The solution is formulated in terms of the test particle's charge-to-mass ratio, energy and angular momentum, as well as the black hole's mass and charge-to-mass ratio.

The rest of this paper is organized as follows. Section \ref{sec:2nd} introduces the harmonic metric of Reissner-Nordstr\"{o}m black hole in the 2PN approximations, the geodesic equation with the Lorentz force, the corresponding Lagrangian, as well as the orbital energy and angular momentum. In Section \ref{sec:3rd} we present the detail derivation on the 2PN solution for the charged test particle's quasi-Keplerian motion. Summary is given in section \ref{sec:4th}.

\section{The 2PN Lagrangian, energy, angular momentum Reissner-Nordstr\"{o}m spacetime}\label{sec:2nd}

The Reissner-Nordstr\"{o}m black hole's mass and electric charge are denoted by $M$ and $Q$. In the harmonic coordinates, the metric of Reissner-Nordstr\"{o}m black hole in the 2PN approximation can be written as~\cite{LinJiang2014}
\begin{eqnarray}
&& g_{00}= -1 +\frac{2M}{r} -\frac{2 M^2}{r^2}\Big(1+\frac{1}{2}\epsilon_{0}^2\Big)+\frac{2M^3}{r^3}(1+\epsilon_{0}^2)~,\label{eq:metric-1nd}\\
&& g_{0i}= 0~,\label{eq:metric-2nd}\\
&& g_{ij}=\Big(1\!+\!\frac{2 M}{r}+\frac{M^2}{r^2}  \Big)\delta_{ij}+\frac{M^2}{r^2}\Big(1-\frac{1}{2}\epsilon_{0}^2\Big)\frac{x^ix^j}{r^2} ,\label{eq:metric-3nd}
\end{eqnarray}
where $\epsilon_{0}\!\equiv\!Q/M$ is the charge-to-mass of the gravitational source. The non naked singularity of the Reissner-Nordstr\"{o}m spacetime requires $|\epsilon_{0}|\!\le\!1$. $r\!\equiv\!|\bm{x}|$ denotes the distance from the field position $\bm{x}\!\equiv\!(x,\,y,\,z)$ to the black hole located at the coordinate origin. The gravitational constant and the speed of light in vacuum are set as $1$ ($G=1$ and $c=1$). The metric has signature of ($-+++$). Latin indices $i$ and $j$ run from 1 to 3.

The charged test particle has mass $m$ and electric charge $q$. The particle's charge-to-mass ratio is $\epsilon_{1}\!\equiv\!q/m$, and its motion is described by the geodesic equation with the Lorentz force
\begin{eqnarray}
&& \frac{d^2x^{\mu}}{d\tau^2}+\Gamma^{\mu}_{\nu\lambda}\frac{dx^{\nu}}{d\tau}\frac{dx^{\lambda}}{d\tau}=\epsilon_{1}{F^{\mu}}_{\nu}\frac{dx^{\nu}}{d\tau} ~,\label{eq:geo}
\end{eqnarray}
where $\Gamma^{\mu}_{\nu\lambda}$ denotes the Christoffel's symbols that given by the derivatives of the chosen metric $g_{\mu\nu}$, $\tau$ is the proper time of the particle along its world line. The electromagnetic Faraday tensor $F_{\mu\nu}$ is given by
\begin{eqnarray}
&& F_{\mu\nu}\!=\!\partial A_{\nu}/\partial x^{\mu}\!-\!\partial A_{\mu}/\partial x^{\nu}~,
\end{eqnarray}
where $A_{\alpha}$ is the associated electromagnetic potential vector
\begin{eqnarray}
&& A_0=-\frac{\epsilon_{0}M}{r}\Big(1+\frac{M}{r}\Big)^{-1}~,\\
&& A_i = 0~.
\end{eqnarray}

Substituting the 2PN metric into Eq.~\eqref{eq:geo}, we can obtain the 2PN equation of motion for the charged test particle as follow
\begin{eqnarray}
&& \frac{d \bm{v}}{dt}=-\frac{M \bm{x}}{r^3}\Big[(1\!-\!\epsilon_{0}\epsilon_{1})\!-\!\frac{M}{r}(4\!+\!\epsilon_{0}^2\!-\!5\epsilon_{0}\epsilon_{1})\!+\!\bm{v}^{2}\Big(1\!+\!\frac{1}{2}\epsilon_{0}\epsilon_{1}\Big)
\!+\!\frac{M^2}{r^2}\Big(9\!+\!6\epsilon_{0}^2\!-\!\frac{27}{2}\epsilon_{0}\epsilon_{1}\!-\!\frac{3}{2}\epsilon_{0}^3\epsilon_{1}\Big)\nn\\
&&\hskip 2.5cm -\frac{2M(\bm{v}\!\cdot\!\bm{x})^2}{r^3}(1\!-\!\epsilon_{0}^2)\!-\!\frac{\bm{v}^{2}}{2}(\epsilon_{0}\epsilon_{1}\!+\!\epsilon_{0}^2)\frac{M}{r}+\frac{1}{8}\bm{v}^{4}\epsilon_{0}\epsilon_{1}\Big] \nn\\
&&\hskip 2.5cm  +\frac{M(\bm{v}\!\cdot\!\bm{x})\bm{v}}{r^3}\Big[(4\!-\!\epsilon_{0}\epsilon_{1}) \!-\!\frac{M}{r}(2\!+\!2\epsilon_{0}^2\!-\!\epsilon_{0}\epsilon_{1})\!+\!\frac{1}{2}\bm{v}^{2}\epsilon_{0}\epsilon_{1}\Big]~,\label{eq:equation-of-motion}
\end{eqnarray}
where $\bm{v}$ denotes the the particle's velocity. When $\epsilon_{1}=0$ this equation reduces to the 2PN dynamics of a neutral particle in Reissner-Nordstr\"{o}m spacetime.

From the equation-of-motion Eq.~(\ref{eq:equation-of-motion}) and the Euler-Lagrange equation
\begin{eqnarray}
&& \frac{d}{dt}\frac{\partial \rm{L}}{\partial\bm{v}}= \frac{\partial \rm{L}}{\partial\bm{x}}~,
\end{eqnarray}
we can obtain the corresponding 2PN Lagrangian of the charged test particle
\begin{eqnarray}
&& {\rm L} = \frac{1}{2}\bm{v}^{2}+\frac{M}{r}(1-\epsilon_{0}\epsilon_{1})+\frac{1}{8}\bm{v}^{4}
+\frac{3}{2}\frac{M}{r}\bm{v}^{2}-\frac{1}{2}\frac{M^2}{r^2}(1+\epsilon_{0}^2-2\epsilon_{0}\epsilon_{1})\!
+\!\frac{1}{2}\frac{M^3}{r^3}(1+\epsilon_{0}^2-2\epsilon_{0}\epsilon_{1})\nn\\
&&~~~~~~+\frac{1}{16}\bm{v}^{6}+\frac{1}{4}\frac{M^2}{r^2}\bm{v}^{2}(7-\epsilon_{0}^2)+\frac{7}{8}\frac{M}{r}\bm{v}^{4}
+\frac{1}{2}\frac{M^2}{r^2}\frac{(\bm{v}\!\cdot\!\bm{x})^2}{r^2}(1\!-\!\epsilon_{0}^2)~,\label{eq:Lagrangian}
\end{eqnarray}
Based on this Lagrangian, we can calculate the 2PN energy $\mathcal{E}$ and angular momentum $\mathcal{J}$ of the charged test particle as follows:
\begin{eqnarray}
&& \mathcal {E}
=\frac{1}{2}\bm{v}^{2}\!-\!\frac{M}{r}(1-\epsilon_{0}\epsilon_{1})\!+\!\frac{3}{8}\bm{v}^{4}\!+\!\frac{3}{2}\frac{M}{r}\bm{v}^{2}\!\!+
\!\frac{1}{2}\frac{M^2}{r^2}(1\!+\!\epsilon_{0}^2-2\epsilon_{0}\epsilon_{1})\!
-\!\frac{1}{2}\frac{M^3}{r^3}(1\!+\!\epsilon_{0}^2-2\epsilon_{0}\epsilon_{1})\!\nn\\
&& ~~~~~~+\frac{5}{16}\bm{v}^{6}\!+\!
\frac{1}{4}\frac{M^2}{r^2}\bm{v}^{2}(7\!-\!\epsilon_{0}^2)\!+\!\frac{21}{8}\frac{M}{r}\bm{v}^{4}
\!+\!\frac{1}{2}\frac{M^2}{r^2}\frac{(\bm{v}\!\cdot\!\bm{x})^2}{r^2}(1\!-\!\epsilon_{0}^2)~,\label{eq:total-energy}\\
&& \mathcal{J}
= |\bm{x}\!\times\!\bm{v}|\Big[1\!+\!\frac{1}{2}\bm{v}^{2}
\!+\!\frac{3M}{r}\!+\!\frac{3}{8}\bm{v}^{4}\!+\!\frac{1}{2}\frac{M^2}{r^2}(7\!-\!\epsilon_{0}^2)\!+\!\frac{7}{2}\frac{M}{r}\bm{v}^{2}\Big]~.
\label{eq:total-angular momentum}
\end{eqnarray}
Notice that the mass $m$ of the test particle has been absorbed in the Lagrangian, the orbital energy and angular momentum.

\section{The quasi-Keplerian motion in the 2PN approximations}\label{sec:3rd}

We follow the same procedure given by Soffel et al.~\cite{Soffel1987} to derive the 2PN solution for the quasi-Keplerian motion of the charged test particle in the Reissner-Nordstr\"{o}m spacetime.

Due to the spherically-symmetry of the Reissner-Nordstr\"{o}m spacetime, we only need to consider the motion of the charged test particle's motion in the equatorial plane, in which the particle's trajectory can be expressed as
\begin{equation}
\bm{x} = r(\cos\phi\,\bm{e}_{x}+\sin\phi\,\bm{e}_{y})~,\label{eq:x1}
\end{equation}
where $\phi$ is the azimuthal angle. $\bm{e}_{x}$ and $\bm{e}_{y}$ are the unit vectors of the $x$-axis and $y$-axis.

The expressions for the orbital energy and angular momentum in Eqs.~(\ref{eq:total-energy})-(\ref{eq:total-angular momentum}) can be written as:
\begin{eqnarray}
&& \mathcal {E}=
\frac{1}{2}(\dot{r}^2\!+\!r^2\dot{\phi}^2)\!-\!\frac{M}{r}(1-\epsilon_{0}\epsilon_{1})\!
+\!\frac{3}{8}(\dot{r}^2\!+\!r^2\dot{\phi}^2)^2\!+\!\frac{3}{2}\frac{M}{r}(\dot{r}^2\!+\!r^2\dot{\phi}^2)
\!+\!\frac{M^2}{2r^2}(1\!+\!\epsilon_{0}^2-2\epsilon_{0}\epsilon_{1})\Big(1\!-\!\frac{M}{r}\Big)\nn\\
&& \hskip 1cm +\frac{5}{16}(\dot{r}^2\!+\!r^2\dot{\phi}^2)^3\!+\!\frac{1}{4}\frac{M^2}{r^2}(\dot{r}^2\!+\!r^2\dot{\phi}^2)(7\!-\!\epsilon_{0}^2)
\!+\!\frac{21}{8}\frac{M}{r}(\dot{r}^2\!+\!r^2\dot{\phi}^2)^2\!+\!\frac{M^2}{2r^2}\dot{r}^2(1\!-\!\epsilon_{0}^2) ,\label{eq:Brumberg-total-energy}\\
&& \hskip -0cm\mathcal{J}^2=r^4\dot{\phi}^2
\Big[1\!+\!\frac{1}{2}(\dot{r}^2\!+\!r^2\dot{\phi}^2)\!+\!\frac{3M}{r}\!+\!\frac{3}{8}(\dot{r}^2\!+\!r^2\dot{\phi}^2)^2\!+\!
\frac{7}{2}\frac{M}{r}(\dot{r}^2\!+\!r^2\dot{\phi}^2)\!+\!\frac{7}{2}\frac{M^2}{r^2}\Big(1\!-\!\frac{1}{7}\epsilon_{0}^2\Big)\Big]^2~,\label{eq:Brumberg-total-angular momentum}
\end{eqnarray}
where the dot denotes the derivative with respect to the time.

From these two expressions we can obtain
\begin{equation}
r^4\dot{\phi}^2=\mathcal{J}^2 \Big\{1\!-\!2\,\mathcal {E}\!-\!\frac{M}{r}(8\!-\!2\epsilon_{0}\epsilon_{1})\!+\!3\,\mathcal {E}^2\!+\!\frac{M^2}{r^2}[34\!-\!18\epsilon_{0}\epsilon_{1}\!+\!\epsilon_{0}^2(2\!+\!3\epsilon_{1}^2)]\!+\!\,\mathcal {E}\frac{M}{r}(16\!-\!6\epsilon_{0}\epsilon_{1})\Big\}~,\label{eq:r4dottheta}
\end{equation}
and
\begin{equation}
 \dot{r}^2=A+\frac{B}{r}+\frac{C}{r^2}+\frac{D}{r^3}+\frac{E}{r^4}~,\label{eq:rdot20}
\end{equation}
with
\begin{eqnarray}
&&\hskip -0.5cm A=2\mathcal {E}\Big(1\!-\!\frac{3}{2}\,\mathcal {E}\!+\!2\,\mathcal {E}^2\Big)~,\\
&&\hskip -0.5cm B=2M[(1\!-\!\epsilon_{0}\epsilon_{1})\!-\!3\,\mathcal {E}(2\!-\!\epsilon_{0}\epsilon_{1})\!+\!3\,\mathcal {E}^2(3-2\epsilon_{0}\epsilon_{1})]~,\\
&&\hskip -0.5cm C=-\mathcal{J}^2\Big\{1\!-\!2\,\mathcal{E}\!+\!\frac{M^2}{\mathcal{J}^2}[10\!-\!14\epsilon_{0}\epsilon_{1}\!+\!\epsilon_{0}^2(1\!+\!3\epsilon_{1}^2)]
\!-\!6\frac{M^2\mathcal {E}}{\mathcal{J}^2}[6\!-\!7\epsilon_{0}\epsilon_{1}\!+\!\epsilon_{0}^2(1\!+\!2\epsilon_{1}^2)]\!+\!3\,\mathcal {E}^2\Big\}~,\\
&&\hskip -0.5cm D=M \mathcal{J}^2\Big\{(8\!-\!2\epsilon_{0}\epsilon_{1})\!-\!\,\mathcal{E}(16\!-\!6\epsilon_{0}\epsilon_{1})\!+\!2\frac{M^2}{\mathcal{J}^2}
[13\!-\!25\epsilon_{0}\epsilon_{1}\!+\!\epsilon_{0}^2(5\!+\!12\epsilon_{1}^2)\!-\!\epsilon_{0}^3\epsilon_{1}(3\!+\!2\epsilon_{1}^2)]\Big\}~,\\
&&\hskip -0.5cm E=-3M^2 \mathcal{J}^2[11\!-\!6\epsilon_{0}\epsilon_{1}\!+\!\epsilon_{0}^2(1\!+\!\epsilon_{1}^2)]~.
\end{eqnarray}

Making use of the relation
\begin{equation}
 \dot{r}^2=\Big[\frac{d(1/r)}{d\phi}\Big]^2(r^4\dot{\phi}^2)~,\label{eq:rdot2-relation}
\end{equation}
and plugging Eqs.~(\ref{eq:r4dottheta})-(\ref{eq:rdot20}) into (\ref{eq:rdot2-relation}), we can write the radial equation in the form
\begin{eqnarray}
&& \Big[\frac{d(1/r)}{d\phi}\Big]^2=A'+\frac{B'}{r}+\frac{C'}{r^2}+\frac{D'}{r^3}+\frac{E'}{r^4}~,\label{eq:the-radial-equation1}
\end{eqnarray}
with
\begin{eqnarray}
&& A'=\frac{2\mathcal {E}}{\mathcal{J}^2}\Big(1\!+\!\frac{1}{2}\mathcal {E}\Big)~,\\
&& B'=\frac{2M}{\mathcal{J}^2}[(1\!-\!\epsilon_{0}\epsilon_{1})\!+\!\,\mathcal {E}(4\!-\!\epsilon_{0}\epsilon_{1})\!+\!2\mathcal {E}^2]~,\\
&& C'=-\Big\{1\!-\!\frac{M^2}{\mathcal{J}^2}[6\!-\!6\epsilon_{0}\epsilon_{1}\!-\!\epsilon_{0}^2(1\!-\!\epsilon_{1}^2)] \!-\!6(2\!-\!\epsilon_{0}\epsilon_{1})\frac{M^2\mathcal {E}}{\mathcal{J}^2}\Big\}~,\\
&& D'=2\frac{M^3}{\mathcal{J}^2}(3\!-\!\epsilon_{0}^2\!-\!3\epsilon_{0}\epsilon_{1}\!+\!\epsilon_{0}^2\epsilon_{1}^2)~,\\
&& E'= M^2(1\!-\!\epsilon_{0}^2)~.
\end{eqnarray}
Since the right hand side of Eq.~(\ref{eq:the-radial-equation1}) is a fourth-order polynomial in $r^{-1}$, we can further re-write it as
\begin{eqnarray}
&& \Big[\frac{d(1/r)}{d\phi}\Big]^2=\Big[\frac{1}{r}\!-\!\frac{1}{a_{r}
(1+e_{r})}\Big]\Big[\frac{1}{a_{r}(1-e_{r})}\!-\!\frac{1}{r}\Big]\Big(C_{1}+\frac{C_{2}}{r}+\frac{C_{3}}{r^2}\Big)~.\label{eq:the-radial-equation2}
\end{eqnarray}
Comparing the coefficients between Eq.~(\ref{eq:the-radial-equation1}) and Eq.~(\ref{eq:the-radial-equation2}), we have
\begin{eqnarray}
&& a_{r}=\frac{M(1\!-\!\epsilon_{0}\epsilon_{1})}{-2\mathcal{E}}\Big\{1\!+\!\frac{1}{2}\mathcal{E}\frac{7\!
-\!\epsilon_{0}\epsilon_{1}}{1\!-\!\epsilon_{0}\epsilon_{1}} \!+\!\frac{1}{4}\mathcal{E}^2\frac{1\!+\!\epsilon_{0}\epsilon_{1}}{1\!-\!\epsilon_{0}\epsilon_{1}}\!+\!2\frac{M^2\mathcal{E}}{\mathcal{J}^2}
\Big[(4\!-\!\epsilon_{0}^2)\!-\!\epsilon_{0}^2
\frac{1\!-\!\epsilon_{1}^2}{1\!-\!\epsilon_{0}\epsilon_{1}}\Big]\Big\}~,\label{eq:aPN_EJ}\\
&& e_r^2=1\!+\!\frac{2\mathcal{E}\mathcal{J}^2}{M^2(1\!-\!\epsilon_{0}\epsilon_{1})^2}\!-\frac{\mathcal{E}}
{(1\!-\!\epsilon_{0}\epsilon_{1})^2}
\Big\{2[6\!-\!6\epsilon_{0}\epsilon_{1}\!-\!\epsilon_{0}^2(1\!-\!\epsilon_{1}^2)]\!+\!
\frac{3\mathcal{E}\mathcal{J}^2}{M^2}\frac{(5\!-\!\epsilon_{0}\epsilon_{1})}{(1\!-\!\epsilon_{0}\epsilon_{1})}\Big\}\nn\\
&& \hskip 1cm +\,\frac{\mathcal{E}^2}{(1\!-\!\epsilon_{0}\epsilon_{1})^2}\Big\{\frac{30\!-\!36\epsilon_{0}\epsilon_{1}\!
-\!3\epsilon_{0}^3\epsilon_{1}(3\!+\!\epsilon_{1}^2)\!+\!\epsilon_{0}^2(5\!+\!13\epsilon_{1}^2)}{
(1\!-\!\epsilon_{0}\epsilon_{1})}\!+\!\frac{2\mathcal{E}\mathcal{J}^2(40\!-\!15\epsilon_{0}\epsilon_{1}\!+\!2\epsilon_{0}^2\epsilon_{1}^2)}
{M^2(1\!-\!\epsilon_{0}\epsilon_{1})^2}\nn\\
&& \hskip 3.55cm  -\frac{8M^2}{\mathcal{E}\mathcal{J}^2}[4\!-\!4\epsilon_{0}\epsilon_{1}\!+\!\epsilon_{0}^3\epsilon_{1}\!-\!\epsilon_{0}^2(2\!-\!\epsilon_{1}^2)]
(1\!-\!\epsilon_{0}\epsilon_{1})\Big\}~,\label{eq:ePN_EJ}\\
&& C_{1}=1\!-\!\frac{M^2}{\mathcal{J}^2}[6\!-\!6\epsilon_{0}\epsilon_{1}\!-\!\epsilon_{0}^2(1\!-\!\epsilon_{1}^2)]\!
-\!2\frac{M^2\mathcal{E}}{\mathcal{J}^2}(7\!-\!\epsilon_{0}^2\!-\!3\epsilon_{0}\epsilon_{1})\nn\\
&& \hskip 1.5cm -\frac{4M^4}{\mathcal{J}^4}
[4\!-\!4\epsilon_{0}\epsilon_{1}\!+\!\epsilon_{0}^3\epsilon_{1}\!-\!\epsilon_{0}^2(2\!-\!\epsilon_{1}^2)](1\!-\!\epsilon_{0}\epsilon_{1})~,\label{eq:C1}\\
&& C_{2}=-\frac{2M^3}{\mathcal{J}^2}[4\!-\!4\epsilon_{0}\epsilon_{1}\!+\!\epsilon_{0}^3\epsilon_{1}\!-\!\epsilon_{0}^2(2\!-\!\epsilon_{1}^2)]~,\label{eq:C2}\\
&& C_{3}=-M^2(1\!-\!\epsilon_{0}^2)~.\label{eq:C3}
\end{eqnarray}
It can be seen from Eq.~(\ref{eq:the-radial-equation2}) that $r_{\pm}=a_{r}(1\pm e_{r})$ represent the maximal and minimal values for $r$. Hence, $a_{r}$ and $e_{r}$ can be regarded as the semi-major axis and the eccentricity of the quasi-Keplerian orbit.

The solution of Eq.~(\ref{eq:the-radial-equation2}) can be written as:
\begin{eqnarray}
&& r=\frac{a_{r}(1-e_{r}^2)}{1+e_{r}\cos{f}}~,\label{eq:rBrumberg}
\end{eqnarray}
with $f$ being the true anomaly for the quasi-Keplerian orbit and obeying
\begin{eqnarray}
&& \Big(\frac{df}{d\phi}\Big)^2=C_{1}+\frac{C_{2}}{r}+\frac{C_{3}}{r^2}~.\label{eq:f_phi2}
\end{eqnarray}

Substituting Eqs.~(\ref{eq:C1})-(\ref{eq:rBrumberg}) into Eq.~(\ref{eq:f_phi2}), we have
\begin{eqnarray}
&& \hskip -0.75cm \frac{df}{d\phi}=F\Big\{1\!-\! \frac{M^4}{\mathcal{J}^4}[5\!-\!8\epsilon_{0}\epsilon_{1}\!-\!\epsilon_{0}^4\epsilon_{1}^2\!-\!\epsilon_{0}^2(3\!-\!5\epsilon_{1}^2)\!
+\!\epsilon_{0}^3\epsilon_{1}(3\!-\!\epsilon_{1}^2)]e_{r} \cos f
\!-\!\frac{M^4}{4\mathcal{J}^4}(1\!-\!\epsilon_{0}^2)e_{r}^2 \cos 2f\Big\},\label{eq:dfdphi}
\end{eqnarray}
with
\begin{eqnarray}
&& \hskip 0cm F=1\!-\!\frac{M^2}{2\mathcal{J}^2}[6\!-\!6\epsilon_{0}\epsilon_{1}\!-\!\epsilon_{0}^2(1\!-\!\epsilon_{1}^2)]\!
-\!\mathcal{E}\frac{M^2}{2\mathcal{J}^2}\Big[\frac{1\!-\!\epsilon_{0}^2}{(1\!-\!\epsilon_{0}\epsilon_{1})^2}\!+\!2(7\!-\!\epsilon_{0}^2\!
-\!3\epsilon_{0}\epsilon_{1})\Big] \nn\\
&& \hskip 1cm -\frac{M^4}{8\mathcal{J}^4}[138\!-\!264\epsilon_{0}\epsilon_{1}\!-\!6\epsilon_{0}^2(11\!-\!28\epsilon_{1}^2)\!+\!\epsilon_{0}^3\epsilon_{1}(84\!-\!36\epsilon_{1}^2)\!+\!\epsilon_{0}^4(1\!-\!26\epsilon_{1}^2
\!+\!\epsilon_{1}^4)]~.\label{eq:F}
\end{eqnarray}

Making integration of Eq.~(\ref{eq:dfdphi}), we can obtain
\begin{equation}
\phi \Big(\frac{2\pi}{\Phi}\Big)=f\!+\!\frac{M^4}{\mathcal{J}^4}[5\!-\!8\epsilon_{0}\epsilon_{1}\!-\!\epsilon_{0}^4\epsilon_{1}^2\!-\!\epsilon_{0}^2(3\!-\!5\epsilon_{1}^2)\!+\!\epsilon_{0}^3\epsilon_{1}
(3\!-\!\epsilon_{1}^2)]e_{r} \sin f\!+\!\frac{M^4}{8\mathcal{J}^4}(1\!-\!\epsilon_{0}^2)e_r^2\sin 2f~,\label{eq:phi_f}
\end{equation}
with
\begin{eqnarray}
&& \hskip 0cm \Phi=2\pi\Big\{1\!+\!\frac{M^2}{2\mathcal{J}^2} [6\!-\!6\epsilon_{0}\epsilon_{1}\!-\!\epsilon_{0}^2(1\!-\!\epsilon_{1}^2)]\!
+\!\mathcal{E}\frac{M^2}{2\mathcal{J}^2}\Big[\frac{1\!-\!\epsilon_{0}^2}{(1\!-\!\epsilon_{0}\epsilon_{1})^2}\!
+\!2(7\!-\!\epsilon_{0}^2\!-\!3\epsilon_{0}\epsilon_{1})\Big] \nn\\
&& \hskip 1cm +\frac{3M^4}{8\mathcal{J}^4}[70\!-\!136\epsilon_{0}\epsilon_{1}\!-\!\epsilon_{0}^2(30\!-\!88\epsilon_{1}^2)\!+\!4\epsilon_{0}^3\epsilon_{1}(9\!-\!5\epsilon_{1}^2)\!+\!\epsilon_{0}^4(1\!-\!10\epsilon_{1}^2
\!+\!\epsilon_{1}^4)]\Big\}~.\label{eq:Phi}
\end{eqnarray}

Finally, we derive the time dependence of the quasi-Keplerian motion.
Combining Eqs.~(\ref{eq:r4dottheta}) and (\ref{eq:dfdphi})-(\ref{eq:F}), we have
\begin{eqnarray}
&& \hskip -0cm r^2\dot{f}=\mathcal{J}\Big\{1\!-\!\mathcal {E}\!-\!\frac{M}{r}(4\!-\!\epsilon_{0}\epsilon_{1})\!-\!\frac{M^2}{2\mathcal{J}^2} [6\!-\!6\epsilon_{0}\epsilon_{1}\!-\!\epsilon_{0}^2(1\!-\!\epsilon_{1}^2)]\!
+\!\mathcal {E}^2\!+\!\frac{M^2}{r^2}[9\!-\!5\epsilon_{0}\epsilon_{1}\!+\!\epsilon_{0}^2(1\!+\!\epsilon_{1}^2)]\nn\\
&& \hskip 1.25cm -\mathcal{E}\frac{M^2}{\mathcal{J}^2}\frac{9\!-\!16\epsilon_{0}\epsilon_{1}\!-\!\epsilon_{0}^2(2\!-\!7\epsilon_{1}^2)\!+\!2\epsilon_{0}^3\epsilon_{1}(1\!+\!\epsilon_{1}^2)\!-\!
\epsilon_{0}^4\epsilon_{1}^2(1\!+\!\epsilon_{1}^2)}{2(1\!-\!\epsilon_{0}\epsilon_{1})^2} \nn\\
&& \hskip 1.25cm +\frac{M}{r}\frac{M^2}{2\mathcal{J}^2}[6\!-\!6\epsilon_{0}\epsilon_{1}\!-\!\epsilon_{0}^2(1\!-\!\epsilon_{1}^2)](4\!-\!\epsilon_{0}\epsilon_{1})\!+\!2\mathcal {E}\frac{M}{r}(2\!-\!\epsilon_{0}\epsilon_{1})\nn\\
&& \hskip 1.25cm -\frac{M^4}{8\mathcal{J}^4}[138\!-\!264\epsilon_{0}\epsilon_{1}\!-\!6\epsilon_{0}^2(11\!-\!28\epsilon_{1}^2)\!+\!\epsilon_{0}^3\epsilon_{1}(84\!-\!36\epsilon_{1}^2)\!+\!\epsilon_{0}^4(1\!-\!26\epsilon_{1}^2
\!+\!\epsilon_{1}^4)] \nn\\
&& \hskip 1.25cm
- \frac{M^4}{\mathcal{J}^4}[5\!-\!8\epsilon_{0}\epsilon_{1}\!-\!\epsilon_{0}^4\epsilon_{1}^2\!-\!\epsilon_{0}^2(3\!-\!5\epsilon_{1}^2)\!+\!\epsilon_{0}^3\epsilon_{1}(3\!-\!\epsilon_{1}^2)]e_{r} \cos f
\!-\!\frac{M^4}{4\mathcal{J}^4}(1\!-\!\epsilon_{0}^2)e_{r}^2 \cos 2f\Big\}.\label{eq:dfdt}
\end{eqnarray}
\indent Introducing the post-Newtonian eccentric anomaly $u$ by the relations
\begin{equation}
\hskip -0cm \sin f\!=\!\frac{(1\!-\!e_{r}^2)^{\frac{1}{2}}\sin u}{1\!-\!e_{r}\cos u};\,\,\cos f\!=\!\frac{\cos u\!-\!e_{r}}{1\!-\!e_{r}\cos u};\,\,
 f\!=\!2 \!\arctan\!\Big(\sqrt{\frac{1\!+\!e_{r}}{1\!-\!e_{r}}}\tan \frac{u}{2}\Big),
\label{eq:u-relation}
\end{equation}
we have
\begin{eqnarray}
\frac{df}{dt}=\frac{(1-e_{r}^2)^{1/2}}{1-e_{r}\cos u}\frac{du}{dt}~,\label{eq:dfdu}
\end{eqnarray}
and we can formulate the orbit given in Eq.~(\ref{eq:rBrumberg}) in terms of $u$ as
\begin{equation}
 r=a_{r}(1-e_{r} \cos u)~.\label{eq:r}
\end{equation}

Integrating Eq.~(\ref{eq:dfdt}) and making use of Eqs.~(\ref{eq:u-relation})-(\ref{eq:r}), we can achieve the final piece of the 2PN solution for the motion in Reissner-Nordstr\"{o}m spacetime.
\begin{equation}
t\Big(\frac{2\pi}{{\rm T}_u}\Big)  = u-e_t \sin{u}+ \frac{2M \,\mathcal{E}^2 }{\sqrt{-2\,\mathcal{E}\mathcal{J}^2}}\frac{1\!-\!\epsilon_{0}^2\!+\!2(1\!-\!\epsilon_{0}\epsilon_{1})^2(7\!-\!\epsilon_{0}^2\!-\!3\epsilon_{0}\epsilon_{1})}{(1\!-\!\epsilon_{0}\epsilon_{1})^3}
(f-u)~,\label{eq:nt}
\end{equation}
with ${\rm T}_{\!u}$ being the period for the eccentric anomaly $u$ of the quasi-Keplerian motion
\begin{eqnarray}
&& {\rm T}_u=\frac{2\pi M(1\!-\!\epsilon_{0}\epsilon_{1})}{(-2\mathcal{E})^{\frac{3}{2}}}\Big[1\!-\!\frac{3}{4}\mathcal{E}\frac{(5\!-\!\epsilon_{0}\epsilon_{1})}{(1\!-\!\epsilon_{0}\epsilon_{1})}\!
-\!\frac{15}{32}\mathcal{E}^2\frac{(7\!+\!\epsilon_{0}\epsilon_{1})}{(1\!-\!\epsilon_{0}\epsilon_{1})} \nn\\
&& \hskip 3cm+\frac{2M \,\mathcal{E}^2 }{\sqrt{-2\,\mathcal{E}\mathcal{J}^2}}\frac{1\!-\!\epsilon_{0}^2\!+\!2(1\!-\!\epsilon_{0}\epsilon_{1})^2
(7\!-\!\epsilon_{0}^2\!-\!3\epsilon_{0}\epsilon_{1})}{(1\!-\!\epsilon_{0}\epsilon_{1})^3}\Big]~,
\end{eqnarray}
and $e_t$ being the time eccentricity
\begin{eqnarray}
&& e_t=e_r\Big[1+2\,\mathcal{E}\frac{(4\!-\!\epsilon_{0}\epsilon_{1})}{(1\!-\!\epsilon_{0}\epsilon_{1})}+\mathcal{E}^2\frac{36\!-\!19\epsilon_{0}\epsilon_{1}\!+\!\epsilon_{0}^2\epsilon_{1}^2}
{(1\!-\!\epsilon_{0}\epsilon_{1})^2}+
2\mathcal{E}\frac{M^2}{\mathcal{J}^2}\frac{4(1\!-\!\epsilon_{0}\epsilon_{1})\!-\!\epsilon_{0}^2(2\!-\!\epsilon_{1}^2\!-\!\epsilon_{0}\epsilon_{1})}{(1\!-\!\epsilon_{0}\epsilon_{1})}\nn\\
&&\hskip 1.75cm
-\frac{2M \,\mathcal{E}^2 }{\sqrt{-2\,\mathcal{E}\mathcal{J}^2}}\frac{1\!-\!\epsilon_{0}^2\!+\!2(1\!-\!\epsilon_{0}\epsilon_{1})^2(7\!-\!\epsilon_{0}^2\!-\!3\epsilon_{0}\epsilon_{1})}{(1\!-\!\epsilon_{0}\epsilon_{1})^3}\Big]~.
\end{eqnarray}
\indent In the literatures, one usually uses another true anomaly $\upsilon$ to replace the true anomaly $f$ in the formula of the quasi-Keplerian equation, requiring that the $\sin \upsilon$ contribution in $\phi (\frac{2\pi}{\Phi})$ vanish at each PN order~\cite{MGS2004,KonigsdorfferGopakumar2005,THS2010}.
Following the same method given in Ref~\cite{THS2010}, we set
\begin{equation}
 \upsilon=2 \arctan \Big(\sqrt{\frac{1+e_{\phi}}{1-e_{\phi}}}\tan \frac{u}{2}\Big)~,\label{eq:upsilon_u}
\end{equation}
with
\begin{equation}
 e_\phi=e_r(1+\epsilon\, c_1+\epsilon^2\, c_2)~,
\end{equation}
differing from the radial eccentricity $e_r$ by some 1PN and 2PN level corrections $c_1$ and $c_2$. Here $\epsilon$ only denotes the PN order and does not have any value. Eliminating $u$ in Eq.~(\ref{eq:u-relation}) with the help of Eq.~(\ref{eq:upsilon_u}), we have~\cite{THS2010}
\begin{equation}
\hskip -0.05cm f\!=\!\upsilon+\epsilon\, c_1\frac{e_r}{e_r^2\!-\!1}\sin\upsilon
+\epsilon^2\,\Big[\Big(c_2\!-\!c_1^2\frac{e_r^2}{e_r^2\!-\!1}\Big)\frac{e_r}{e_r^2
\!-\!1}\sin \upsilon+\frac{c_1^2}{4}\frac{e_r^2}{(e_r^2\!-\!1)^2}\sin 2\upsilon\Big]. 
\end{equation}
Substituting this result into Eq.~(\ref{eq:phi_f}) and requiring the $\sin \upsilon$ term to vanish in $\phi (\frac{2\pi}{\Phi})$, we can obtain
\begin{eqnarray}
&& c_1=0~,\\
&& c_2=-\,2\,\mathcal{E}\,\frac{M^2}{\mathcal{J}^2}\,\frac{5\!-\!8\epsilon_{0}\epsilon_{1}\!-\!\epsilon_{0}^2(3\!-\!5\epsilon_{1}^2\!+\!\epsilon_{0}^2\epsilon_{1}^2)\!+\!\epsilon_{0}^3\epsilon_{1}(3\!-\!\epsilon_{1}^2)}
{(1\!-\!\epsilon_{0}\epsilon_{1})^2}~,
\end{eqnarray}
which lead to
\begin{equation}
e_{\phi} = e_r \Big[1\!-\!2\,\mathcal{E}\,\frac{M^2}{\mathcal{J}^2}\,\frac{5\!-\!8\epsilon_{0}\epsilon_{1}\!-\!\epsilon_{0}^2(3\!-\!5\epsilon_{1}^2\!+\!\epsilon_{0}^2\epsilon_{1}^2)\!+\!\epsilon_{0}^3\epsilon_{1}(3\!-\!\epsilon_{1}^2)}
{(1\!-\!\epsilon_{0}\epsilon_{1})^2}\Big]~,
\end{equation}
\begin{equation}
 \phi \Big(\frac{2\pi}{\Phi}\Big)= \upsilon +\frac{M^4}{8\mathcal{J}^4}(1\!-\!\epsilon_{0}^2)\Big[1\!+\!\frac{2\mathcal{E}\mathcal{J}^2}{M^2(1\!-\!\epsilon_{0}\epsilon_{1})^2}\Big]\sin 2\upsilon~.
\end{equation}
\indent With the true anomaly $\upsilon$, we can re-express the time dependance of the quasi-Keplerian motion Eq.~(\ref{eq:nt}) in the form of
\begin{equation}
 t\Big(\frac{2\pi}{{\rm T}_u}\Big)  = u-e_t \sin{u}+ \frac{2M \,\mathcal{E}^2 }{\sqrt{-2\,\mathcal{E}\mathcal{J}^2}}\frac{1\!-\!\epsilon_{0}^2\!+\!2(1\!-\!\epsilon_{0}\epsilon_{1})^2(7\!-\!\epsilon_{0}^2\!-\!3\epsilon_{0}\epsilon_{1})}{(1\!-\!\epsilon_{0}\epsilon_{1})^3} (\upsilon-u)~.
\end{equation}
\indent Notice that $|\epsilon_0| \!\le\! 1$ and $|\epsilon_0\epsilon_1|\!\ll\! 1$ are assumed in the above derivations, and all the formulas are valid up to the 2PN accuracy.

\section{Summary}\label{sec:4th}

Basing on the 2PN metric of the Reissner-Nordstr\"{o}m black hole in the harmonic coordinates and the geodesic equation with the Lorentz force, we first calculate the corresponding Lagrangian, orbital energy and angular momentum of the charged test particle. Then, through a function fitting method, we obtain the orbital parameters. Finally, we derive the quasi-Keplerian equation of the charged test particle in the Reissner-Nordstr\"{o}m spacetime. We obtain two slightly different but equivalent formulations in the 2PN approximations. The results are summarized as follows.

The first formulation can be expressed as
\begin{eqnarray}
&& \bm{x} = r(\cos\phi\,\bm{e}_{x}+\sin\phi\,\bm{e}_{y})~,\nn\\
&& r=a_{r}(1-e_{r} \cos u)~,\nn\\
&& \phi \Big(\frac{2\pi}{\Phi}\Big)=f+ N_0\sin f+N_1\sin 2f~,\nn\\
&& f=2 \arctan \Big(\sqrt{\frac{1+e_{r}}{1-e_{r}}}\tan \frac{u}{2}\Big)~,\nn\\
&&  t\Big(\frac{2\pi}{{\rm T}_u}\Big)  = u-e_t \sin{u}+ N_2 (f-u)~,\nn
\end{eqnarray}
and the second formulation can be expressed as
\begin{eqnarray}
&& \bm{x} = r(\cos\phi\,\bm{e}_{x}+\sin\phi\,\bm{e}_{y})~,\nn\\
&& r=a_{r}(1-e_{r} \cos u)~,\nn\\
&& \phi \Big(\frac{2\pi}{\Phi}\Big)= \upsilon + N_1 \sin 2\upsilon~,\nn\\
&& \upsilon=2 \arctan \Big(\sqrt{\frac{1+e_{\phi}}{1-e_{\phi}}}\tan \frac{u}{2}\Big)~,\nn\\
&&  t\Big(\frac{2\pi}{{\rm T}_u}\Big)  = u-e_t \sin{u}+ N_2 (\upsilon-u)~,\nn
\end{eqnarray}
where
\begin{eqnarray}
&& a_{r}=\frac{M(1\!-\!\epsilon_{0}\epsilon_{1})}{-2\mathcal{E}}\Big\{1\!+\!\frac{1}{2}\mathcal{E}\frac{7\!-\!\epsilon_{0}\epsilon_{1}}{1\!-\!\epsilon_{0}\epsilon_{1}} \!+\!\frac{1}{4}\mathcal{E}^2\frac{1\!+\!\epsilon_{0}\epsilon_{1}}{1\!-\!\epsilon_{0}\epsilon_{1}}\!+\!2\frac{M^2\mathcal{E}}{\mathcal{J}^2}\Big[(4\!-\!\epsilon_{0}^2)\!-\!\epsilon_{0}^2
\frac{1\!-\!\epsilon_{1}^2}{1\!-\!\epsilon_{0}\epsilon_{1}}\Big]\Big\}~,\nn\\
&& e_r^2=1\!+\!\frac{2\mathcal{E}\mathcal{J}^2}{M^2(1\!-\!\epsilon_{0}\epsilon_{1})^2}\!-\frac{\mathcal{E}}{(1\!-\!\epsilon_{0}\epsilon_{1})^2}
\Big\{2[6\!-\!6\epsilon_{0}\epsilon_{1}\!-\!\epsilon_{0}^2(1\!-\!\epsilon_{1}^2)]\!+\!
\frac{3\mathcal{E}\mathcal{J}^2}{M^2}\frac{(5\!-\!\epsilon_{0}\epsilon_{1})}{(1\!-\!\epsilon_{0}\epsilon_{1})}\Big\}\nn\\
&& \hskip 1cm+\frac{\mathcal{E}^2}{(1\!-\!\epsilon_{0}\epsilon_{1})^2}\Big\{\frac{30\!+\!5\epsilon_{0}^2\!-\!36\epsilon_{0}\epsilon_{1}\!-\!3\epsilon_{0}^3\epsilon_{1}(3\!+\!\epsilon_{1}^2)\!
+\!13\epsilon_{0}^2\epsilon_{1}^2}{
(1\!-\!\epsilon_{0}\epsilon_{1})}\!+\!\frac{2\mathcal{E}\mathcal{J}^2(40\!-\!15\epsilon_{0}\epsilon_{1}\!+\!2\epsilon_{0}^2\epsilon_{1}^2)}{M^2(1\!-\!\epsilon_{0}\epsilon_{1})^2} \nn\\
&& \hskip 3.45cm  -\frac{8M^2}{\mathcal{E}\mathcal{J}^2}[4\!-\!4\epsilon_{0}\epsilon_{1}\!+\!\epsilon_{0}^3\epsilon_{1}\!-\!\epsilon_{0}^2(2\!-\!\epsilon_{1}^2)](1\!-\!\epsilon_{0}\epsilon_{1})\Big\}~,\nn\\
&&  e_t=e_r\Big[1+2\,\mathcal{E}\frac{(4\!-\!\epsilon_{0}\epsilon_{1})}{(1\!-\!\epsilon_{0}\epsilon_{1})}+\mathcal{E}^2\frac{36\!-\!19\epsilon_{0}\epsilon_{1}\!+\!\epsilon_{0}^2\epsilon_{1}^2}
{(1\!-\!\epsilon_{0}\epsilon_{1})^2}+
2\mathcal{E}\frac{M^2}{\mathcal{J}^2}\frac{4(1\!-\!\epsilon_{0}\epsilon_{1})\!-\!\epsilon_{0}^2(2\!-\!\epsilon_{1}^2\!-\!\epsilon_{0}\epsilon_{1})}{(1\!-\!\epsilon_{0}\epsilon_{1})}\nn\\
&&\hskip 1.5cm
-\frac{2M \,\mathcal{E}^2 }{\sqrt{-2\,\mathcal{E}\mathcal{J}^2}}\frac{1\!-\!\epsilon_{0}^2\!+\!2(1\!-\!\epsilon_{0}\epsilon_{1})^2(7\!-\!\epsilon_{0}^2\!-\!3\epsilon_{0}\epsilon_{1})}{(1\!-\!\epsilon_{0}\epsilon_{1})^3}\Big]~,\nn\\
&& e_{\phi} = e_r \Big[1\!-\!2\,\mathcal{E}\,\frac{M^2}{\mathcal{J}^2}\,\frac{5\!-\!8\epsilon_{0}\epsilon_{1}\!-\!\epsilon_{0}^2(3\!-\!5\epsilon_{1}^2\!+\!\epsilon_{0}^2\epsilon_{1}^2)\!+\!\epsilon_{0}^3\epsilon_{1}(3\!-\!\epsilon_{1}^2)}
{(1\!-\!\epsilon_{0}\epsilon_{1})^2}\Big]~,\nn\\
&&\Phi=2\pi\Big\{1\!+\!\frac{M^2}{2\mathcal{J}^2} [6(1\!-\!\epsilon_{0}\epsilon_{1})\!-\!\epsilon_{0}^2(1\!-\!\epsilon_{1}^2)]\!
+\!\mathcal{E}\frac{M^2}{2\mathcal{J}^2}\Big[\frac{1\!-\!\epsilon_{0}^2}{(1\!-\!\epsilon_{0}\epsilon_{1})^2}\!+\!2(7\!-\!\epsilon_{0}^2\!-\!3\epsilon_{0}\epsilon_{1})\Big] \nn\\
&& \hskip 1.75cm +\frac{3M^4}{8\mathcal{J}^4}[70\!-\!136\epsilon_{0}\epsilon_{1}\!-\!\epsilon_{0}^2(30\!-\!88\epsilon_{1}^2)\!+\!4\epsilon_{0}^3\epsilon_{1}(9\!-\!5\epsilon_{1}^2)\!+\!\epsilon_{0}^4(1\!-\!10\epsilon_{1}^2
\!+\!\epsilon_{1}^4)]\Big\}~,\nn\\
&&  N_0= \frac{M^4}{\mathcal{J}^4}[5\!-\!8\epsilon_{0}\epsilon_{1}\!-\!\epsilon_{0}^4\epsilon_{1}^2\!-\!\epsilon_{0}^2(3\!-\!5\epsilon_{1}^2)\!+\!\epsilon_{0}^3\epsilon_{1}
(3\!-\!\epsilon_{1}^2)] \Big[1\!+\!\frac{2\mathcal{E}\mathcal{J}^2}{M^2(1\!-\!\epsilon_{0}\epsilon_{1})^2}\Big]^{\frac{1}{2}}~,\nn\\
&&  N_1= \frac{M^4}{8\mathcal{J}^4}(1\!-\!\epsilon_{0}^2)\Big[1\!+\!\frac{2\mathcal{E}\mathcal{J}^2}{M^2(1\!-\!\epsilon_{0}\epsilon_{1})^2}\Big]~,\nn\\
&& N_2 =\frac{2M \,\mathcal{E}^2 }{\sqrt{-2\,\mathcal{E}\mathcal{J}^2}}\frac{1\!-\!\epsilon_{0}^2\!+\!2(1\!-\!\epsilon_{0}\epsilon_{1})^2(7\!-\!\epsilon_{0}^2\!-\!3\epsilon_{0}\epsilon_{1})}{(1\!-\!\epsilon_{0}\epsilon_{1})^3}~,\nn\\
&& {\rm T}_u=\frac{2\pi M(1\!-\!\epsilon_{0}\epsilon_{1})}{(-2\mathcal{E})^{\frac{3}{2}}}\Big[1\!-\!\frac{3}{4}\mathcal{E}\frac{(5\!-\!\epsilon_{0}\epsilon_{1})}{(1\!-\!\epsilon_{0}\epsilon_{1})}\!
-\!\frac{15}{32}\mathcal{E}^2\frac{(7\!+\!\epsilon_{0}\epsilon_{1})}{(1\!-\!\epsilon_{0}\epsilon_{1})} \nn\\
&& \hskip 3cm+\frac{2M \,\mathcal{E}^2 }{\sqrt{-2\,\mathcal{E}\mathcal{J}^2}}\frac{1\!-\!\epsilon_{0}^2\!+\!2(1\!-\!\epsilon_{0}\epsilon_{1})^2
(7\!-\!\epsilon_{0}^2\!-\!3\epsilon_{0}\epsilon_{1})}{(1\!-\!\epsilon_{0}\epsilon_{1})^3}\Big]~.\nn
\end{eqnarray}
In the formulations, $a_{r}$, $e_{r}$ and $u$ can be regarded as the semi-major axis, the eccentricity, the eccentric anomaly of the quasi-Keplerian motion in the post-Newtonian approximations. $f$ and $\upsilon$ are two slightly different definitions of the true anomaly. ${\rm T}_u$ denotes the orbital period. The difference between $\Phi$ and $2\pi$ is the perihelion precession. The effects of the black hole's charge on the test particle's motion including perihelion precession and orbital period are characterized by the terms containing $\epsilon_0$, and the effects of the test particle's charge are described by the terms containing $\epsilon_1$.
The achieved 2PN solution can be applied to the motion of the electrically charged test particles with small charge-to-mass ratio in the Reissner-Nordstr\"{o}m spacetime which has $|\epsilon_0| \! \le \! 1$ for the non naked singularity, e.g., the typically charged solar mass object in the field of the charged supermassive black hole, and also to the motion of the test particle with arbitrary charge-to-mass ratio in the field of the weakly charged black hole as long as $|\epsilon_0\epsilon_1 |\!\ll\! 1$.

\section*{ACKNOWLEDGEMENT}
We thank the referee for providing constructive suggestions to promote the quality of this paper. This work was supported in part by the National Natural Science Foundation of China (Grant Nos. 11973025 and 12147208).

\bibliography{Reference_20200707}

\begin{thebibliography}{43}%
\makeatletter
\providecommand \@ifxundefined [1]{%
 \@ifx{#1\undefined}
}%
\providecommand \@ifnum [1]{%
 \ifnum #1\expandafter \@firstoftwo
 \else \expandafter \@secondoftwo
 \fi
}%
\providecommand \@ifx [1]{%
 \ifx #1\expandafter \@firstoftwo
 \else \expandafter \@secondoftwo
 \fi
}%
\providecommand \natexlab [1]{#1}%
\providecommand \enquote  [1]{``#1''}%
\providecommand \bibnamefont  [1]{#1}%
\providecommand \bibfnamefont [1]{#1}%
\providecommand \citenamefont [1]{#1}%
\providecommand \href@noop [0]{\@secondoftwo}%
\providecommand \href [0]{\begingroup \@sanitize@url \@href}%
\providecommand \@href[1]{\@@startlink{#1}\@@href}%
\providecommand \@@href[1]{\endgroup#1\@@endlink}%
\providecommand \@sanitize@url [0]{\catcode `\\12\catcode `\$12\catcode
  `\&12\catcode `\#12\catcode `\^12\catcode `\_12\catcode `\%12\relax}%
\providecommand \@@startlink[1]{}%
\providecommand \@@endlink[0]{}%
\providecommand \url  [0]{\begingroup\@sanitize@url \@url }%
\providecommand \@url [1]{\endgroup\@href {#1}{\urlprefix }}%
\providecommand \urlprefix  [0]{URL }%
\providecommand \Eprint [0]{\href }%
\providecommand \doibase [0]{http://dx.doi.org/}%
\providecommand \selectlanguage [0]{\@gobble}%
\providecommand \bibinfo  [0]{\@secondoftwo}%
\providecommand \bibfield  [0]{\@secondoftwo}%
\providecommand \translation [1]{[#1]}%
\providecommand \BibitemOpen [0]{}%
\providecommand \bibitemStop [0]{}%
\providecommand \bibitemNoStop [0]{.\EOS\space}%
\providecommand \EOS [0]{\spacefactor3000\relax}%
\providecommand \BibitemShut  [1]{\csname bibitem#1\endcsname}%
\let\auto@bib@innerbib\@empty
\bibitem [{\citenamefont {Brumberg}(1972)}]{Brumberg1972}%
  \BibitemOpen
  \bibfield  {author} {\bibinfo {author} {\bibfnamefont {V.}~\bibnamefont
  {Brumberg}},\ }\href@noop {} {\emph {\bibinfo {title} {Relativistic
  celesctial mechanics}}}\ (\bibinfo  {publisher} {Nauka, Moscow in Russian},\
  \bibinfo {year} {1972})\BibitemShut {NoStop}%
\bibitem [{\citenamefont {Soffel}\ \emph {et~al.}(1987)\citenamefont {Soffel},
  \citenamefont {Ruder},\ and\ \citenamefont {Schneider}}]{Soffel1987}%
  \BibitemOpen
  \bibfield  {author} {\bibinfo {author} {\bibfnamefont {M.~H.}\ \bibnamefont
  {Soffel}}, \bibinfo {author} {\bibfnamefont {H.}~\bibnamefont {Ruder}}, \
  and\ \bibinfo {author} {\bibfnamefont {M.}~\bibnamefont {Schneider}},\
  }\href@noop {} {\bibfield  {journal} {\bibinfo  {journal} {Celestial Mech.}\
  }\textbf {\bibinfo {volume} {40}},\ \bibinfo {pages} {77} (\bibinfo {year}
  {1987})}\BibitemShut {NoStop}%
\bibitem [{\citenamefont {Soffel}(1989)}]{Soffel1989}%
  \BibitemOpen
  \bibfield  {author} {\bibinfo {author} {\bibfnamefont {M.~H.}\ \bibnamefont
  {Soffel}},\ }\href@noop {} {\emph {\bibinfo {title} {Relativity in
  Astrometry, Celestial Mechanics and Geodesy}}}\ (\bibinfo  {publisher}
  {Berlin: Springer},\ \bibinfo {year} {1989})\BibitemShut {NoStop}%
\bibitem [{\citenamefont {Klioner}\ and\ \citenamefont
  {Kopeikin}(1994)}]{KlionerKopeikin1994}%
  \BibitemOpen
  \bibfield  {author} {\bibinfo {author} {\bibfnamefont {S.~A.}\ \bibnamefont
  {Klioner}}\ and\ \bibinfo {author} {\bibfnamefont {S.~M.}\ \bibnamefont
  {Kopeikin}},\ }\href@noop {} {\bibfield  {journal} {\bibinfo  {journal}
  {Astrophys. J.}\ }\textbf {\bibinfo {volume} {427}},\ \bibinfo {pages} {951}
  (\bibinfo {year} {1994})}\BibitemShut {NoStop}%
\bibitem [{\citenamefont {Kopeikin}\ \emph {et~al.}(2012)\citenamefont
  {Kopeikin}, \citenamefont {Efroimsky},\ and\ \citenamefont
  {Kaplan}}]{KopeikinEfroimskyKaplan2012}%
  \BibitemOpen
  \bibfield  {author} {\bibinfo {author} {\bibfnamefont {S.~M.}\ \bibnamefont
  {Kopeikin}}, \bibinfo {author} {\bibfnamefont {M.}~\bibnamefont {Efroimsky}},
  \ and\ \bibinfo {author} {\bibfnamefont {G.}~\bibnamefont {Kaplan}},\
  }\href@noop {} {\emph {\bibinfo {title} {Relativistic Celestical Mechanics of
  the Solar System}}}\ (\bibinfo  {publisher} {Wiley-VCH},\ \bibinfo {address}
  {New York},\ \bibinfo {year} {2012})\BibitemShut {NoStop}%
\bibitem [{\citenamefont {Damour}\ and\ \citenamefont
  {Sch\"{a}fer}(1988)}]{DamourSchafer1988}%
  \BibitemOpen
  \bibfield  {author} {\bibinfo {author} {\bibfnamefont {T.}~\bibnamefont
  {Damour}}\ and\ \bibinfo {author} {\bibfnamefont {G.}~\bibnamefont
  {Sch\"{a}fer}},\ }\href@noop {} {\bibfield  {journal} {\bibinfo  {journal}
  {Nuovo Cimento B}\ }\textbf {\bibinfo {volume} {101}},\ \bibinfo {pages}
  {127} (\bibinfo {year} {1988})}\BibitemShut {NoStop}%
\bibitem [{\citenamefont {Sch\"{a}fer}\ and\ \citenamefont
  {Wex}(1993)}]{SchaferWex1993}%
  \BibitemOpen
  \bibfield  {author} {\bibinfo {author} {\bibfnamefont {G.}~\bibnamefont
  {Sch\"{a}fer}}\ and\ \bibinfo {author} {\bibfnamefont {N.}~\bibnamefont
  {Wex}},\ }\href@noop {} {\bibfield  {journal} {\bibinfo  {journal} {Phys.
  Lett. A}\ }\textbf {\bibinfo {volume} {174}},\ \bibinfo {pages} {196}
  (\bibinfo {year} {1993})}\BibitemShut {NoStop}%
\bibitem [{\citenamefont {Memmesheimer}\ \emph {et~al.}(2004)\citenamefont
  {Memmesheimer}, \citenamefont {Gopakumar},\ and\ \citenamefont
  {Sch\"{a}fer}}]{MGS2004}%
  \BibitemOpen
  \bibfield  {author} {\bibinfo {author} {\bibfnamefont {R.~M.}\ \bibnamefont
  {Memmesheimer}}, \bibinfo {author} {\bibfnamefont {A.}~\bibnamefont
  {Gopakumar}}, \ and\ \bibinfo {author} {\bibfnamefont {G.}~\bibnamefont
  {Sch\"{a}fer}},\ }\href@noop {} {\bibfield  {journal} {\bibinfo  {journal}
  {Phys. Rev. D}\ }\textbf {\bibinfo {volume} {70}},\ \bibinfo {pages} {104011}
  (\bibinfo {year} {2004})}\BibitemShut {NoStop}%
\bibitem [{\citenamefont {Boetzel}\ \emph {et~al.}(2017)\citenamefont
  {Boetzel}, \citenamefont {Susobhanan}, \citenamefont {Gopakumar},
  \citenamefont {Klein},\ and\ \citenamefont {Jetzer}}]{Boetzel2017}%
  \BibitemOpen
  \bibfield  {author} {\bibinfo {author} {\bibfnamefont {Y.}~\bibnamefont
  {Boetzel}}, \bibinfo {author} {\bibfnamefont {A.}~\bibnamefont {Susobhanan}},
  \bibinfo {author} {\bibfnamefont {A.}~\bibnamefont {Gopakumar}}, \bibinfo
  {author} {\bibfnamefont {A.}~\bibnamefont {Klein}}, \ and\ \bibinfo {author}
  {\bibfnamefont {P.}~\bibnamefont {Jetzer}},\ }\href@noop {} {\bibfield
  {journal} {\bibinfo  {journal} {Phys. Rev. D}\ }\textbf {\bibinfo {volume}
  {96}},\ \bibinfo {pages} {044011} (\bibinfo {year} {2017})}\BibitemShut
  {NoStop}%
\bibitem [{\citenamefont {Cho}\ \emph {et~al.}(2018)\citenamefont {Cho},
  \citenamefont {Gopakumar}, \citenamefont {Haney},\ and\ \citenamefont
  {Lee}}]{Cho2018}%
  \BibitemOpen
  \bibfield  {author} {\bibinfo {author} {\bibfnamefont {G.}~\bibnamefont
  {Cho}}, \bibinfo {author} {\bibfnamefont {A.}~\bibnamefont {Gopakumar}},
  \bibinfo {author} {\bibfnamefont {M.}~\bibnamefont {Haney}}, \ and\ \bibinfo
  {author} {\bibfnamefont {H.~M.}\ \bibnamefont {Lee}},\ }\href@noop {}
  {\bibfield  {journal} {\bibinfo  {journal} {Phys. Rev. D}\ }\textbf {\bibinfo
  {volume} {98}},\ \bibinfo {pages} {024039} (\bibinfo {year}
  {2018})}\BibitemShut {NoStop}%
\bibitem [{\citenamefont {Soffel}\ and\ \citenamefont
  {Han}(2019)}]{SoffelHan2019}%
  \BibitemOpen
  \bibfield  {author} {\bibinfo {author} {\bibfnamefont {M.~H.}\ \bibnamefont
  {Soffel}}\ and\ \bibinfo {author} {\bibfnamefont {W.~B.}\ \bibnamefont
  {Han}},\ }\href@noop {} {\emph {\bibinfo {title} {Applied General Relativity:
  Theory and Applications in Astronomy, Celestial Mechanics and Metrology}}}\
  (\bibinfo  {publisher} {Springer},\ \bibinfo {year} {2019})\BibitemShut
  {NoStop}%
\bibitem [{\citenamefont {Wex}(1995)}]{Wex1995}%
  \BibitemOpen
  \bibfield  {author} {\bibinfo {author} {\bibfnamefont {N.}~\bibnamefont
  {Wex}},\ }\href@noop {} {\bibfield  {journal} {\bibinfo  {journal} {Class.
  Quantum Gravity}\ }\textbf {\bibinfo {volume} {12}},\ \bibinfo {pages} {983}
  (\bibinfo {year} {1995})}\BibitemShut {NoStop}%
\bibitem [{\citenamefont {Gergely}\ \emph
  {et~al.}(1998{\natexlab{a}})\citenamefont {Gergely}, \citenamefont
  {Perj\'{e}s},\ and\ \citenamefont {Vas\'{u}th}}]{GPV1998a}%
  \BibitemOpen
  \bibfield  {author} {\bibinfo {author} {\bibfnamefont {L.~A.}\ \bibnamefont
  {Gergely}}, \bibinfo {author} {\bibfnamefont {Z.~I.}\ \bibnamefont
  {Perj\'{e}s}}, \ and\ \bibinfo {author} {\bibfnamefont {M.}~\bibnamefont
  {Vas\'{u}th}},\ }\href@noop {} {\bibfield  {journal} {\bibinfo  {journal}
  {Phys. Rev. D}\ }\textbf {\bibinfo {volume} {57}},\ \bibinfo {pages} {876}
  (\bibinfo {year} {1998}{\natexlab{a}})}\BibitemShut {NoStop}%
\bibitem [{\citenamefont {Gergely}\ \emph
  {et~al.}(1998{\natexlab{b}})\citenamefont {Gergely}, \citenamefont
  {Perj\'{e}s},\ and\ \citenamefont {Vas\'{u}th}}]{GPV1998b}%
  \BibitemOpen
  \bibfield  {author} {\bibinfo {author} {\bibfnamefont {L.~A.}\ \bibnamefont
  {Gergely}}, \bibinfo {author} {\bibfnamefont {Z.~I.}\ \bibnamefont
  {Perj\'{e}s}}, \ and\ \bibinfo {author} {\bibfnamefont {M.}~\bibnamefont
  {Vas\'{u}th}},\ }\href@noop {} {\bibfield  {journal} {\bibinfo  {journal}
  {Phys. Rev. D}\ }\textbf {\bibinfo {volume} {57}},\ \bibinfo {pages} {3423}
  (\bibinfo {year} {1998}{\natexlab{b}})}\BibitemShut {NoStop}%
\bibitem [{\citenamefont {Gergely}\ \emph
  {et~al.}(1998{\natexlab{c}})\citenamefont {Gergely}, \citenamefont
  {Perj\'{e}s},\ and\ \citenamefont {Vas\'{u}th}}]{GPV1998c}%
  \BibitemOpen
  \bibfield  {author} {\bibinfo {author} {\bibfnamefont {L.~A.}\ \bibnamefont
  {Gergely}}, \bibinfo {author} {\bibfnamefont {Z.~I.}\ \bibnamefont
  {Perj\'{e}s}}, \ and\ \bibinfo {author} {\bibfnamefont {M.}~\bibnamefont
  {Vas\'{u}th}},\ }\href@noop {} {\bibfield  {journal} {\bibinfo  {journal}
  {Phys. Rev. D}\ }\textbf {\bibinfo {volume} {58}},\ \bibinfo {pages} {124001}
  (\bibinfo {year} {1998}{\natexlab{c}})}\BibitemShut {NoStop}%
\bibitem [{\citenamefont {K$\ddot{o}$nigsd$\ddot{o}$rffer}\ and\ \citenamefont
  {Gopakumar}(2005)}]{KonigsdorfferGopakumar2005}%
  \BibitemOpen
  \bibfield  {author} {\bibinfo {author} {\bibfnamefont {C.}~\bibnamefont
  {K$\ddot{o}$nigsd$\ddot{o}$rffer}}\ and\ \bibinfo {author} {\bibfnamefont
  {A.}~\bibnamefont {Gopakumar}},\ }\href@noop {} {\bibfield  {journal}
  {\bibinfo  {journal} {Phys. Rev. D}\ }\textbf {\bibinfo {volume} {71}},\
  \bibinfo {pages} {024039} (\bibinfo {year} {2005})}\BibitemShut {NoStop}%
\bibitem [{\citenamefont {Keresztes}\ \emph {et~al.}(2005)\citenamefont
  {Keresztes}, \citenamefont {Mik\'{o}czi},\ and\ \citenamefont
  {Gergely}}]{KMG2005}%
  \BibitemOpen
  \bibfield  {author} {\bibinfo {author} {\bibfnamefont {Z.}~\bibnamefont
  {Keresztes}}, \bibinfo {author} {\bibfnamefont {B.}~\bibnamefont
  {Mik\'{o}czi}}, \ and\ \bibinfo {author} {\bibfnamefont {L.~A.}\ \bibnamefont
  {Gergely}},\ }\href@noop {} {\bibfield  {journal} {\bibinfo  {journal} {Phys.
  Rev. D}\ }\textbf {\bibinfo {volume} {72}},\ \bibinfo {pages} {104022}
  (\bibinfo {year} {2005})}\BibitemShut {NoStop}%
\bibitem [{\citenamefont {K$\ddot{o}$nigsd$\ddot{o}$rffer}\ and\ \citenamefont
  {Gopakumar}(2006)}]{KonigsdorfferGopakumar2006b}%
  \BibitemOpen
  \bibfield  {author} {\bibinfo {author} {\bibfnamefont {C.}~\bibnamefont
  {K$\ddot{o}$nigsd$\ddot{o}$rffer}}\ and\ \bibinfo {author} {\bibfnamefont
  {A.}~\bibnamefont {Gopakumar}},\ }\href@noop {} {\bibfield  {journal}
  {\bibinfo  {journal} {Phys. Rev. D}\ }\textbf {\bibinfo {volume} {73}},\
  \bibinfo {pages} {044011} (\bibinfo {year} {2006})}\BibitemShut {NoStop}%
\bibitem [{\citenamefont {Gopakumar}\ and\ \citenamefont
  {Sch\"{a}fer}(2011)}]{GopakumarSchafer2011}%
  \BibitemOpen
  \bibfield  {author} {\bibinfo {author} {\bibfnamefont {A.}~\bibnamefont
  {Gopakumar}}\ and\ \bibinfo {author} {\bibfnamefont {G.}~\bibnamefont
  {Sch\"{a}fer}},\ }\href@noop {} {\bibfield  {journal} {\bibinfo  {journal}
  {Phys. Rev. D}\ }\textbf {\bibinfo {volume} {84}},\ \bibinfo {pages} {124007}
  (\bibinfo {year} {2011})}\BibitemShut {NoStop}%
\bibitem [{\citenamefont {Boh\'{e}}\ \emph {et~al.}(2013)\citenamefont
  {Boh\'{e}}, \citenamefont {Marsat}, \citenamefont {Faye},\ and\ \citenamefont
  {Blanchet}}]{BMFB2013}%
  \BibitemOpen
  \bibfield  {author} {\bibinfo {author} {\bibfnamefont {A.}~\bibnamefont
  {Boh\'{e}}}, \bibinfo {author} {\bibfnamefont {S.}~\bibnamefont {Marsat}},
  \bibinfo {author} {\bibfnamefont {G.}~\bibnamefont {Faye}}, \ and\ \bibinfo
  {author} {\bibfnamefont {L.}~\bibnamefont {Blanchet}},\ }\href@noop {}
  {\bibfield  {journal} {\bibinfo  {journal} {Class. Quantum Gravity}\ }\textbf
  {\bibinfo {volume} {30}},\ \bibinfo {pages} {075017} (\bibinfo {year}
  {2013})}\BibitemShut {NoStop}%
\bibitem [{\citenamefont {Gergely}\ and\ \citenamefont
  {Keresztes}(2015)}]{GergelyKeresztes2015}%
  \BibitemOpen
  \bibfield  {author} {\bibinfo {author} {\bibfnamefont {L.~A.}\ \bibnamefont
  {Gergely}}\ and\ \bibinfo {author} {\bibfnamefont {Z.}~\bibnamefont
  {Keresztes}},\ }\href@noop {} {\bibfield  {journal} {\bibinfo  {journal}
  {Phys. Rev. D}\ }\textbf {\bibinfo {volume} {91}},\ \bibinfo {pages} {024012}
  (\bibinfo {year} {2015})}\BibitemShut {NoStop}%
\bibitem [{\citenamefont {Mik\'{o}czi}(2017)}]{Mikoczi2017}%
  \BibitemOpen
  \bibfield  {author} {\bibinfo {author} {\bibfnamefont {B.}~\bibnamefont
  {Mik\'{o}czi}},\ }\href@noop {} {\bibfield  {journal} {\bibinfo  {journal}
  {Phys. Rev. D}\ }\textbf {\bibinfo {volume} {95}},\ \bibinfo {pages} {064023}
  (\bibinfo {year} {2017})}\BibitemShut {NoStop}%
\bibitem [{\citenamefont {Poisson}(1998)}]{Poisson1998}%
  \BibitemOpen
  \bibfield  {author} {\bibinfo {author} {\bibfnamefont {E.}~\bibnamefont
  {Poisson}},\ }\href@noop {} {\bibfield  {journal} {\bibinfo  {journal} {Phys.
  Rev. D}\ }\textbf {\bibinfo {volume} {57}},\ \bibinfo {pages} {5287}
  (\bibinfo {year} {1998})}\BibitemShut {NoStop}%
\bibitem [{\citenamefont {Yang}\ and\ \citenamefont
  {Lin}(2020{\natexlab{a}})}]{YangLin2020a}%
  \BibitemOpen
  \bibfield  {author} {\bibinfo {author} {\bibfnamefont {B.}~\bibnamefont
  {Yang}}\ and\ \bibinfo {author} {\bibfnamefont {W.}~\bibnamefont {Lin}},\
  }\href@noop {} {\bibfield  {journal} {\bibinfo  {journal} {Gen. Relativ.
  Gravit.}\ }\textbf {\bibinfo {volume} {52}},\ \bibinfo {pages} {49} (\bibinfo
  {year} {2020}{\natexlab{a}})}\BibitemShut {NoStop}%
\bibitem [{\citenamefont {Yang}\ and\ \citenamefont
  {Lin}(2020{\natexlab{b}})}]{YangLin2020b}%
  \BibitemOpen
  \bibfield  {author} {\bibinfo {author} {\bibfnamefont {B.}~\bibnamefont
  {Yang}}\ and\ \bibinfo {author} {\bibfnamefont {W.}~\bibnamefont {Lin}},\
  }\href@noop {} {\bibfield  {journal} {\bibinfo  {journal} {Eur. Phys. J.
  Plus}\ }\textbf {\bibinfo {volume} {135}},\ \bibinfo {pages} {137} (\bibinfo
  {year} {2020}{\natexlab{b}})}\BibitemShut {NoStop}%
\bibitem [{\citenamefont {Yang}\ and\ \citenamefont
  {Lin}(2020{\natexlab{c}})}]{YangLin2020d}%
  \BibitemOpen
  \bibfield  {author} {\bibinfo {author} {\bibfnamefont {B.}~\bibnamefont
  {Yang}}\ and\ \bibinfo {author} {\bibfnamefont {W.}~\bibnamefont {Lin}},\
  }\href@noop {} {\bibfield  {journal} {\bibinfo  {journal} {Gravit. Cosmo.}\
  }\textbf {\bibinfo {volume} {26}},\ \bibinfo {pages} {373} (\bibinfo {year}
  {2020}{\natexlab{c}})}\BibitemShut {NoStop}%
\bibitem [{\citenamefont {Yang}\ and\ \citenamefont
  {Lin}(2020{\natexlab{d}})}]{YangLin2020c}%
  \BibitemOpen
  \bibfield  {author} {\bibinfo {author} {\bibfnamefont {B.}~\bibnamefont
  {Yang}}\ and\ \bibinfo {author} {\bibfnamefont {W.}~\bibnamefont {Lin}},\
  }\href@noop {} {\bibfield  {journal} {\bibinfo  {journal} {Phys. Scr.}\
  }\textbf {\bibinfo {volume} {95}},\ \bibinfo {pages} {105008} (\bibinfo
  {year} {2020}{\natexlab{d}})}\BibitemShut {NoStop}%
\bibitem [{\citenamefont {Yang}\ and\ \citenamefont {Lin}(2021)}]{YangLin2021}%
  \BibitemOpen
  \bibfield  {author} {\bibinfo {author} {\bibfnamefont {B.}~\bibnamefont
  {Yang}}\ and\ \bibinfo {author} {\bibfnamefont {W.}~\bibnamefont {Lin}},\
  }\href@noop {} {\bibfield  {journal} {\bibinfo  {journal} {Phys. Scr.}\
  }\textbf {\bibinfo {volume} {96}},\ \bibinfo {pages} {085007} (\bibinfo
  {year} {2021})}\BibitemShut {NoStop}%
\bibitem [{\citenamefont {Reissner}(1916)}]{Reissner1916}%
  \BibitemOpen
  \bibfield  {author} {\bibinfo {author} {\bibfnamefont {H.}~\bibnamefont
  {Reissner}},\ }\href@noop {} {\bibfield  {journal} {\bibinfo  {journal}
  {Annalen Der Physik}\ }\textbf {\bibinfo {volume} {355}},\ \bibinfo {pages}
  {106} (\bibinfo {year} {1916})}\BibitemShut {NoStop}%
\bibitem [{\citenamefont {Weyl}(1917)}]{Weyl1917}%
  \BibitemOpen
  \bibfield  {author} {\bibinfo {author} {\bibfnamefont {H.}~\bibnamefont
  {Weyl}},\ }\href@noop {} {\bibfield  {journal} {\bibinfo  {journal} {Annalen
  Der Physik}\ }\textbf {\bibinfo {volume} {359}},\ \bibinfo {pages} {117}
  (\bibinfo {year} {1917})}\BibitemShut {NoStop}%
\bibitem [{\citenamefont {Nordstr\"{o}m}(1918)}]{Nordstrom1918}%
  \BibitemOpen
  \bibfield  {author} {\bibinfo {author} {\bibfnamefont {G.}~\bibnamefont
  {Nordstr\"{o}m}},\ }\href@noop {} {\bibfield  {journal} {\bibinfo  {journal}
  {Proc. K. Ned. Akad. Wetensch.}\ }\textbf {\bibinfo {volume} {20}},\ \bibinfo
  {pages} {1238} (\bibinfo {year} {1918})}\BibitemShut {NoStop}%
\bibitem [{\citenamefont {Iorio}(2012)}]{Iorio2012}%
  \BibitemOpen
  \bibfield  {author} {\bibinfo {author} {\bibfnamefont {L.}~\bibnamefont
  {Iorio}},\ }\href@noop {} {\bibfield  {journal} {\bibinfo  {journal} {Gen.
  Relativ. Gravit.}\ }\textbf {\bibinfo {volume} {44}},\ \bibinfo {pages}
  {1753} (\bibinfo {year} {2012})}\BibitemShut {NoStop}%
\bibitem [{\citenamefont {Zhao}\ and\ \citenamefont {Xie}(2016)}]{ZhaoXie2016}%
  \BibitemOpen
  \bibfield  {author} {\bibinfo {author} {\bibfnamefont {S.}~\bibnamefont
  {Zhao}}\ and\ \bibinfo {author} {\bibfnamefont {Y.}~\bibnamefont {Xie}},\
  }\href@noop {} {\bibfield  {journal} {\bibinfo  {journal} {JCAP}\ }\textbf
  {\bibinfo {volume} {07}},\ \bibinfo {pages} {007} (\bibinfo {year}
  {2016})}\BibitemShut {NoStop}%
\bibitem [{\citenamefont {Wang}\ \emph {et~al.}(2019)\citenamefont {Wang},
  \citenamefont {Shen},\ and\ \citenamefont {Xie}}]{WangShenXie2019}%
  \BibitemOpen
  \bibfield  {author} {\bibinfo {author} {\bibfnamefont {C.}~\bibnamefont
  {Wang}}, \bibinfo {author} {\bibfnamefont {Y.}~\bibnamefont {Shen}}, \ and\
  \bibinfo {author} {\bibfnamefont {Y.}~\bibnamefont {Xie}},\ }\href@noop {}
  {\bibfield  {journal} {\bibinfo  {journal} {JCAP}\ }\textbf {\bibinfo
  {volume} {04}},\ \bibinfo {pages} {022} (\bibinfo {year} {2019})}\BibitemShut
  {NoStop}%
\bibitem [{\citenamefont {Bini}\ \emph {et~al.}(2000)\citenamefont {Bini},
  \citenamefont {Geralico},\ and\ \citenamefont
  {Ruffini}}]{BiniGeralicoRuffini2000}%
  \BibitemOpen
  \bibfield  {author} {\bibinfo {author} {\bibfnamefont {D.}~\bibnamefont
  {Bini}}, \bibinfo {author} {\bibfnamefont {A.}~\bibnamefont {Geralico}}, \
  and\ \bibinfo {author} {\bibfnamefont {R.}~\bibnamefont {Ruffini}},\
  }\href@noop {} {\bibfield  {journal} {\bibinfo  {journal} {Phys. Rev. D}\
  }\textbf {\bibinfo {volume} {61}},\ \bibinfo {pages} {064013} (\bibinfo
  {year} {2000})}\BibitemShut {NoStop}%
\bibitem [{\citenamefont {Ali}\ and\ \citenamefont
  {Ahmed}(2000)}]{Hossain2000}%
  \BibitemOpen
  \bibfield  {author} {\bibinfo {author} {\bibfnamefont {M.~H.}\ \bibnamefont
  {Ali}}\ and\ \bibinfo {author} {\bibfnamefont {M.}~\bibnamefont {Ahmed}},\
  }\href@noop {} {\bibfield  {journal} {\bibinfo  {journal} {Annals of
  Physics}\ }\textbf {\bibinfo {volume} {282}},\ \bibinfo {pages} {157}
  (\bibinfo {year} {2000})}\BibitemShut {NoStop}%
\bibitem [{\citenamefont {Bini}\ \emph {et~al.}(2007)\citenamefont {Bini},
  \citenamefont {Geralico},\ and\ \citenamefont
  {Ruffini}}]{BiniGeralicoRuffini2007}%
  \BibitemOpen
  \bibfield  {author} {\bibinfo {author} {\bibfnamefont {D.}~\bibnamefont
  {Bini}}, \bibinfo {author} {\bibfnamefont {A.}~\bibnamefont {Geralico}}, \
  and\ \bibinfo {author} {\bibfnamefont {R.}~\bibnamefont {Ruffini}},\
  }\href@noop {} {\bibfield  {journal} {\bibinfo  {journal} {Physics Letters
  A}\ }\textbf {\bibinfo {volume} {360}},\ \bibinfo {pages} {515} (\bibinfo
  {year} {2007})}\BibitemShut {NoStop}%
\bibitem [{\citenamefont {Pugliese}\ \emph {et~al.}(2011)\citenamefont
  {Pugliese}, \citenamefont {Quevedo},\ and\ \citenamefont
  {Ruffini}}]{PuglieseQuevedoRuffini2011b}%
  \BibitemOpen
  \bibfield  {author} {\bibinfo {author} {\bibfnamefont {D.}~\bibnamefont
  {Pugliese}}, \bibinfo {author} {\bibfnamefont {H.}~\bibnamefont {Quevedo}}, \
  and\ \bibinfo {author} {\bibfnamefont {R.}~\bibnamefont {Ruffini}},\
  }\href@noop {} {\bibfield  {journal} {\bibinfo  {journal} {Physical Review
  D}\ }\textbf {\bibinfo {volume} {83}},\ \bibinfo {pages} {024021} (\bibinfo
  {year} {2011})}\BibitemShut {NoStop}%
\bibitem [{\citenamefont {Pugliese}\ \emph {et~al.}(2017)\citenamefont
  {Pugliese}, \citenamefont {Quevedo},\ and\ \citenamefont
  {Ruffini}}]{PuglieseQuevedoRuffini2017}%
  \BibitemOpen
  \bibfield  {author} {\bibinfo {author} {\bibfnamefont {D.}~\bibnamefont
  {Pugliese}}, \bibinfo {author} {\bibfnamefont {H.}~\bibnamefont {Quevedo}}, \
  and\ \bibinfo {author} {\bibfnamefont {R.}~\bibnamefont {Ruffini}},\
  }\href@noop {} {\bibfield  {journal} {\bibinfo  {journal} {European Physical
  Journal C}\ }\textbf {\bibinfo {volume} {77}},\ \bibinfo {pages} {206}
  (\bibinfo {year} {2017})}\BibitemShut {NoStop}%
\bibitem [{\citenamefont {Yang}\ and\ \citenamefont
  {Lin}(2019)}]{YangLin2019b}%
  \BibitemOpen
  \bibfield  {author} {\bibinfo {author} {\bibfnamefont {B.}~\bibnamefont
  {Yang}}\ and\ \bibinfo {author} {\bibfnamefont {W.}~\bibnamefont {Lin}},\
  }\href@noop {} {\bibfield  {journal} {\bibinfo  {journal} {Gen. Relativ.
  Gravit.}\ }\textbf {\bibinfo {volume} {51}},\ \bibinfo {pages} {116}
  (\bibinfo {year} {2019})}\BibitemShut {NoStop}%
\bibitem [{\citenamefont {Bakry}\ \emph {et~al.}(2021)\citenamefont {Bakry},
  \citenamefont {Moatimid},\ and\ \citenamefont
  {Tantawy}}]{BakryMoatimidTantawy2021}%
  \BibitemOpen
  \bibfield  {author} {\bibinfo {author} {\bibfnamefont {M.~A.}\ \bibnamefont
  {Bakry}}, \bibinfo {author} {\bibfnamefont {G.~M.}\ \bibnamefont {Moatimid}},
  \ and\ \bibinfo {author} {\bibfnamefont {M.~M.}\ \bibnamefont {Tantawy}},\
  }\href@noop {} {\bibfield  {journal} {\bibinfo  {journal} {International
  Journal of Modern Physics A}\ }\textbf {\bibinfo {volume} {36}},\ \bibinfo
  {pages} {2150073} (\bibinfo {year} {2021})}\BibitemShut {NoStop}%
\bibitem [{\citenamefont {Lin}\ and\ \citenamefont
  {Jiang}(2014)}]{LinJiang2014}%
  \BibitemOpen
  \bibfield  {author} {\bibinfo {author} {\bibfnamefont {W.}~\bibnamefont
  {Lin}}\ and\ \bibinfo {author} {\bibfnamefont {C.}~\bibnamefont {Jiang}},\
  }\href@noop {} {\bibfield  {journal} {\bibinfo  {journal} {Phys. Rev. D}\
  }\textbf {\bibinfo {volume} {89}},\ \bibinfo {pages} {087502} (\bibinfo
  {year} {2014})}\BibitemShut {NoStop}%
\bibitem [{\citenamefont {Tessmer}\ \emph {et~al.}(2010)\citenamefont
  {Tessmer}, \citenamefont {Hartung},\ and\ \citenamefont
  {Sch\"{a}fer}}]{THS2010}%
  \BibitemOpen
  \bibfield  {author} {\bibinfo {author} {\bibfnamefont {M.}~\bibnamefont
  {Tessmer}}, \bibinfo {author} {\bibfnamefont {J.}~\bibnamefont {Hartung}}, \
  and\ \bibinfo {author} {\bibfnamefont {G.}~\bibnamefont {Sch\"{a}fer}},\
  }\href@noop {} {\bibfield  {journal} {\bibinfo  {journal} {Class. Quantum
  Gravity}\ }\textbf {\bibinfo {volume} {27}},\ \bibinfo {pages} {165005}
  (\bibinfo {year} {2010})}\BibitemShut {NoStop}%
\end{thebibliography}%

\end{document}